\documentstyle[prd,aps,epsfig]{revtex}
\begin{document}

\title{Quark/gluon content of  $\eta(1295)$ and  $\eta(1440)$}
\author{A. V. Anisovich}
\address{Petersburg Nuclear Physics Institute, Gatchina, 188300,
Russia}
\maketitle

\begin{abstract}

The quark/gluon content of  $\eta(1295)$ and  $\eta(1440)$ mesons is
discussed, mesons being considered as members of the first radial
excitation $2 ^1S_0$ $q \bar q$ nonet. Recent results on
 $\eta(1295)$ and  $\eta(1440)$ two-photon widths from L3 together
with the information on radiative $J/\Psi$ decay allow one to evaluate
the $\eta(1295)/ \eta(1440)$ mixing angle and the admixture
of the glueball component. We found that $\eta(1440)$ is dominantly
non-strange $q \bar q$ state, with a possible admixture of the glueball
component $(20\pm 20)\%$.

\end{abstract}

\section{Introduction}

The investigation of pseudoscalar-isoscalar
states ($IJ^{PC}=00^{-+}$) is an important task for
the  $q \bar q$ nonet classification and search for  exotic states.
In this way the study of nature of
$\eta(1295)$ and $\eta(1440)$ which are considered as  members
of the first radial excitation nonet is a crucial step.

During last decades  $\eta(1440)$ has been considered as a probable
candidate for the glueball. This hypothesis is based on the fact that
$\eta(1440)$ is produced in a gluon-rich reaction like radiative
$J/\Psi$ decay. At present this interpretation does not agree with
lattice QCD calculations \cite{Morningstar,UKQCD} which predict
masses of pure  gluonic $0^{-+}$ state in the region above 2 GeV.

Although there is more sophisticated interpretation of $\eta(1440)$
as a bound state of gluinos \cite{Farrar}, it seems that the
interpretation of $\eta(1440)$ as a member of the first radial
excitation $2^1S_0$ $q \bar q$ nonet is the most preferable. As was
demonstrated in \cite{Systematics}, the majority of $q \bar q$
trajectories on the $(n,M^2)$ and $(J,M^2)$ plots ($n$ is radial
quantum number and $J$ is meson spin) are linear, with nearly the same
slope. In  this scheme $\eta(1295)$ is considered as a flavour partner of
$\eta(1440)$.
 The recent results from GAMS \cite{GAMS} and E852
\cite{E852} Collaborations give a clear confirmation of the existence of
$\eta(1295)$ resonance.

The existence of two $0^{-+}$ resonances near 1440 MeV, namely,
$\eta_L$ and $\eta_H$, needs special discussion. According to
PDG-2000 \cite{PDG} there is now  fairly consistent picture for
two pseudoscalars. It was claimed that $\eta_L$ decays
mainly into $a_0(980)\pi$ or directly to $K \bar K \pi$, while the
second one, $\eta_H$, is seen in $K \bar K \pi$ channel only, coming
from $K^*(892)K$ intermediate state.

It seems that this opinion is due to the circumstance that most analyses
have been done with the Breit-Wigner constant width.
In fact, since the $K^* K$ has the threshold at 1394 MeV,
its phase space increases  rapidly at 1400-1500 MeV,
thus rather strong $s$-dependence of the width in this channel takes place.
As was shown in \cite{BES98}, the data on $J/\Psi \to
\gamma(K^+K^-\pi^0)$ and  $J/\Psi \to \gamma(\eta \pi^+ \pi^-)$ can
be well fitted as only one  Breit-Wigner resonance, with s-dependent
width.  This gives  a mass shift  between $\eta\pi\pi$ and $K
K \pi$ channels  about 40 MeV.  Also the analysis of $p \bar p \to \eta
\pi^+ \pi^- \pi^+ \pi^-$ \cite{Nana} shows an  evidence that the
$\eta\pi\pi$ channel is fed by triangle diagrams of the form
$\eta(1440) \to  K^*\bar K$, followed by $K\bar K$ rescattering via
$a_0(980)  \to \eta\pi$.

Let us also note that there is an evidence
\cite{Bugg99}
for the existence of a broad $0^{-+}$ meson which was seen in
radiative $J/\Psi$ decay and has the mass around 2200 MeV.  It was
suggested in \cite{Bugg2000} that this state displayed exotic
characteristics and can be identified as glueball.
In this case this glueball state,
due to its large width ($\sim 800$ MeV),  can  mix with
$\eta(1440)$. The question is whether this mixture is small or large.

Recently the $e^+ e^- \rightarrow e^+ e^- K^0_S K^{\pm} \pi^{\mp}$ and
$e^+e^- \rightarrow e^+e^-\eta \pi^+ \pi^-$ final states were studied
 \cite{L3} with
 L3 detector at LEP. The performed analysis shows the clear signal from
 $\eta(1440)$ in untagged $\gamma\gamma$ collisions in the $ K^0_S
K^{\pm}\pi ^{\mp}$ decay channel. The value of its two-photon width is
found to be
\begin{eqnarray}
\Gamma_{\gamma\gamma}(\eta(1440)) \times
BR(\eta(1440) \rightarrow K\bar K \pi) \;=\; 212 \pm 50 (\mathrm{stat})
\pm 23 (\mathrm{sys}) \; \mathrm{eV}\;.
\label{1}
\end{eqnarray}
In the $\eta \pi^+\pi^-$ decay channel neither $\eta(1440)$ nor
$\eta(1295)$ were observed, and the following upper limits for their
two-photon widths were determined:
\begin{eqnarray}
\Gamma_{\gamma\gamma}(\eta(1440)) \times BR(\eta(1440)
\rightarrow \eta \pi\pi)
\;<\; 95  \; \mathrm{eV}\;.    \label{2}           \\
\Gamma_{\gamma\gamma}(\eta(1295)) \times BR(\eta(1295)
\rightarrow \eta \pi\pi)
\;<\; 66  \; \mathrm{eV}\;.    \label{3}
\end{eqnarray}

The transition amplitudes $\eta(1295)\to\gamma\gamma$ and
$\eta(1440)\to\gamma\gamma$
were calculated in \cite{AANM} under the assumption that the decaying
mesons are  members of the first radial excitation
$2^1S_0$ $q\bar q$ nonet. The calculations show that
partial widths strongly depend on the mixing angle between
 $n\bar n$ and  $s\bar s$ components of the mesons: the main contribution
is due to the $n\bar n$  meson component, while the
contribution of the $s\bar s$ component is small. Neglecting  a possible
mixture with  glueball state, partial widths are found to be of
the order of 100 eV.

The purpose of the present paper is to enlighten the
quark/gluon content of  $\eta(1295)$ and  $\eta(1440)$ using
the two-photon widths and radiative $J/\Psi$ decays  of
these mesons.

\section{ Partial widths for the decays $\eta(1295)\to\gamma\gamma$
and $\eta(1440)\to\gamma\gamma$}

It is well known that isoscalar member of basic pseudoscalar
octet mixes with corresponding pseudoscalar singlet to produce
$\eta$ and $\eta'$. The same situation is expected for members
of the first radial excitation nonet, $\eta(1295)$ and $\eta(1440)$.
These states  can also mix with the glueball component.

Taking into account the mixing of
non-strange quark component, $n\bar n=(u\bar u+d\bar d)/\sqrt{2}$,
with strange one, $s\bar s$, and with glueball component, $G$,
the wave functions are determined as follows:

\begin{eqnarray}
\Psi_{\eta(1440)}&=&\cos\phi\left[\sin\theta\;\psi_{n\bar n}+
\cos\theta\;\psi_{s\bar s}\right]+\sin\phi\psi_G \ ,
 \\
\Psi_{\eta(1295)}&=&\cos\phi'\;\left[\cos\theta'\;\psi_{n\bar n}-
\sin\theta'\;\psi_{s\bar s}\right]+\sin\phi'\;\psi_G\ ,
\label{etawfs}
\end{eqnarray}
where $\theta$ and $\theta'$ are  mixing angles between non-strange
and strange components of the wave function,  and $\phi$, $\phi'$
determine the admixture of the glueball state.
The orthogonality condition reads
\begin{eqnarray}
\cos\phi\cos\phi'\ \sin(\theta'-\theta)+\sin\phi\sin\phi'=0 \ .
\label{ortcon}
\end{eqnarray}

The two-photon decays of $\eta$ and $\eta'$ were analysed in
\cite{AMN}. The data on
 transition form factors $\pi^0\to\gamma^*(Q^2)\gamma$,
$\eta\to\gamma^*(Q^2)\gamma$, $\eta'\to\gamma^*(Q^2)\gamma$ over a broad
range of photon virtualities, $Q^2\le 20$ GeV$^2$ \cite{exper},
 made it possible:\\
(i) to restore wave functions of $\eta$ and $\eta'$ (for both
$n\bar n$ and $s\bar s$ components),\\
(ii) to estimate gluonium admixture in $\eta$ and
$\eta'$,\\
(iii) to restore vertex function for the transition $\gamma\to
q\bar q$ (or photon wave function) as a function of the $q\bar q$
invariant mass.

\subsection{Wave functions and transition form factors}

Both non-strange and strange components of the $\eta(M)$-meson
wave functions  are parametrized in  the exponential form. For the basic and
first radial excitation nonets, the wave functions
are determined as follows:
\begin{equation}
\label{wfparam}
\Psi^{(0)}_\eta(s)=Ce^{-bs}, \quad
\Psi^{(1)}_\eta(s)=C_1(D_1 s-1)e^{-b_1 s},
\end{equation}
where $s$ is the $q \bar q$ invariant mass squared.
Parameters $b$ and $b_1$ are related to the radii squared of
corresponding $\eta(M)$-mesons. Then the other constants ($C$, $C_1$,
$D_1$) are fixed by the normalization and orthogonality conditions:
\begin{equation}
\label{wfparam1}
\Psi^{(0)}_\eta\otimes\Psi^{(0)}_\eta=1, \quad
\Psi^{(1)}_\eta\otimes\Psi^{(1)}_\eta=1, \quad
\Psi^{(0)}_\eta\otimes\Psi^{(1)}_\eta=0.
\end{equation}
The convolution of the $\eta(M)$-meson
 wave function determines the form factor of $\eta$-meson,
$f^{(n)}_\eta(q^2_\perp)=
\left[\Psi^{(n)}_\eta\otimes\Psi^{(n)}_\eta\right]$,
thus allowing us to relate the parameter $b$ (or $b_1$)
at small $Q^2$  to  the $\eta$-meson radius
squared: $f_\eta(q^2_\perp)\simeq 1-\frac 16 R^2_\eta Q^2$.

\begin{figure}[h!]
\begin{tabular}{cc}
\mbox{\epsfig{file=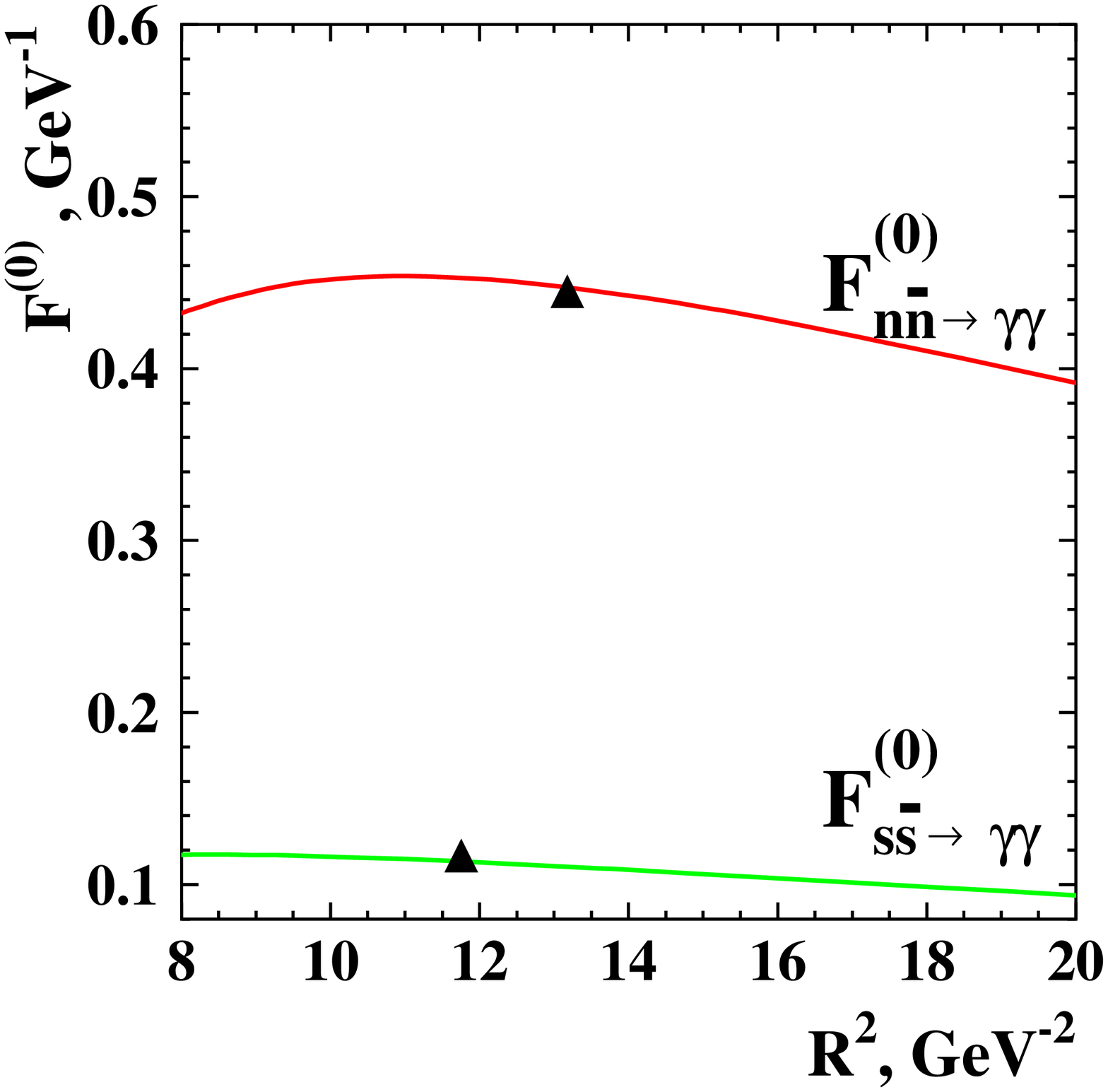,width=0.45\textwidth}} &
\mbox{\epsfig{file=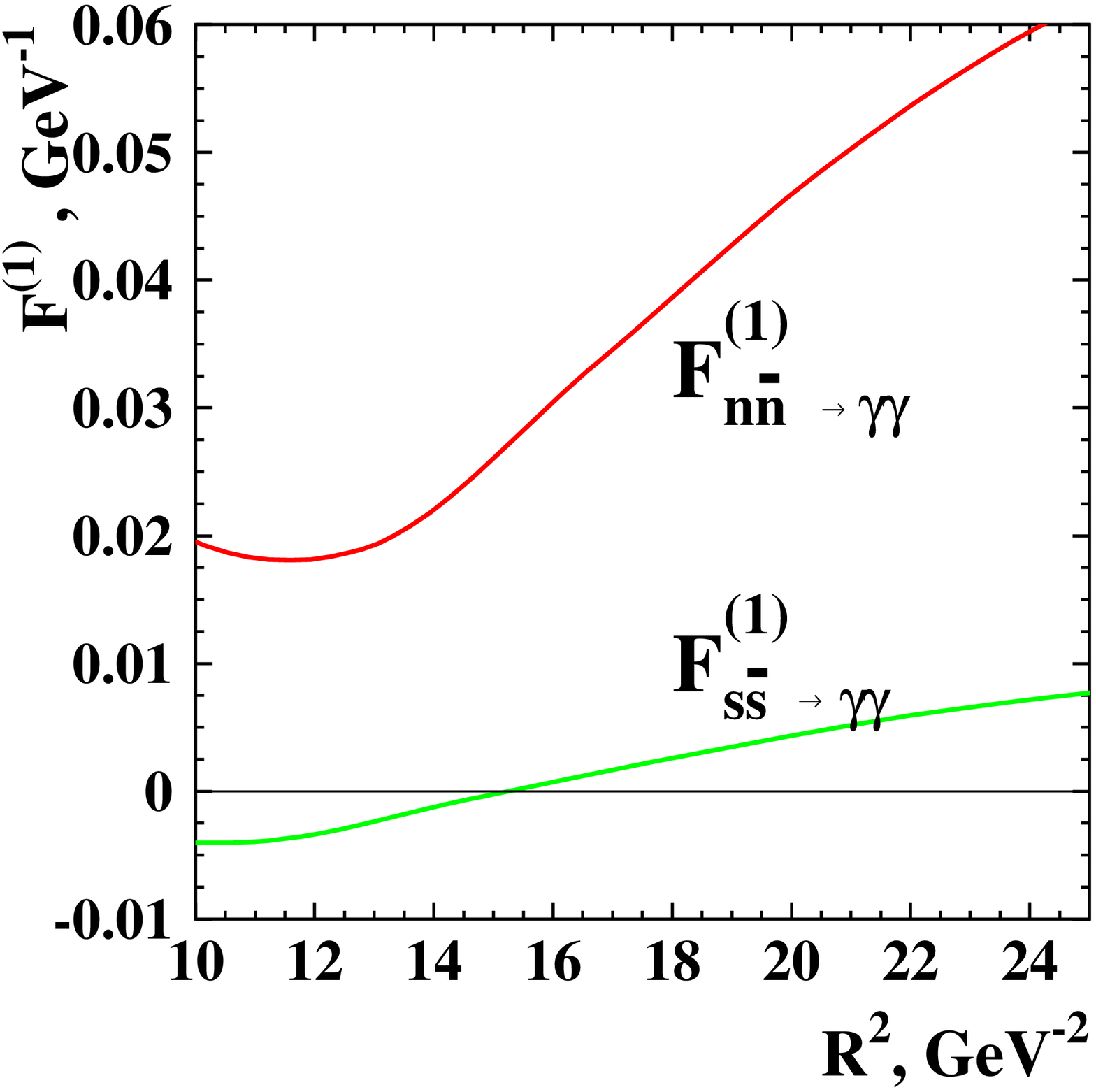,width=0.45\textwidth}}
\end{tabular}
    \caption{\small \it
(a) Transition form factors as function of mean radius squared of
$\eta(M)$ for the basic nonet. Solid curves present calculations with
exponential parametrization (\ref{wfparam}); the rhombuses give
values calculated with the wave function found in [18].
(b) The transition form factors  as function of mean radius squared of
$\eta(M)$ for the first radial-excitation nonet.
}
\label{fig2}
\end{figure}

It is useful to compare the results of calculation of the transition
form factor approximated by simple exponential wave function
with those obtained using  more
sophisticated wave function. Such a comparison can be
done for basic $1^1S_0$ $q\bar q$ nonet.
The  $\eta$ and $\eta'$ wave functions (or those for its
$n\bar n$ and $s\bar s$
components) were found in \cite{AMN} basing on the data
for the transitions $\eta\to\gamma\gamma^*(Q^2)$,
$\eta\to\gamma\gamma^*(Q^2)$ at $Q^2\le 20$ GeV$^2$.
The calculated decay form factors
$F^{(0)}_{n\bar n\to\gamma\gamma}$ and
$F^{(0)}_{s\bar s\to\gamma\gamma}$ at $Q^2=0$ for these wave
functions are marked in Fig. \ref{fig2}a by rhombuses. The wave functions found
in Ref. \cite{AMN} give the following mean radii squared for $n\bar n$
and $s\bar s$ components: $R^2_{n\bar n}=13.1$ GeV$^{-2}$ and
$R^2_{s\bar s}=11.7$ GeV$^{-2}$.
Solid curves in Fig.  \ref{fig2}a represent
$F^{(0)}_{n\bar n\to\gamma\gamma}$ and
$F^{(0)}_{s\bar s\to\gamma\gamma}$ calculated by using
exponential parametrisation (\ref{wfparam}): we see that both
calculations coincide with each other within reasonable accuracy.
The coincidence of the results justifies
exponential approximation for the wave function used in the calculation of
transition form factors at $Q^2 \sim 0$.

The Fig. \ref{fig2}b  shows the  calculation results obtained in
\cite{AANM} for transition form factors  of the first
radial excitation nonet at $Q^2=0$. The form factor of the
$n\bar n$ component,  $F^{(1)}_{n\bar n\to\gamma\gamma}$,
depends strongly on the mean radius squared,
increasing rapidly for $R^2_{n\bar n}\sim 14-24$ GeV$^{-2}$
(0.7-1.2 fm$^2$). As to $s\bar s$ component, the form factor
$F^{(1)}_{s\bar s\to\gamma\gamma}$ is small and it changes sign at
$R^2_{s\bar s}\simeq 15$ GeV$^{-2}$.

The results shown in Fig.\ref{fig2} need to be commented. The dominant
contribution from the non-strange component of the $\eta$-meson wave function
comes mainly from different charge factors of the transition form factors.
For case of $F^{(n)}_{n\bar n\to\gamma\gamma}$ this factor is equal to
 $(q_u^2 + q_d^2)/\sqrt 2$, while for $F^{(n)}_{s\bar s\to\gamma\gamma}$ it is
$q_s^2$, where $q_u, q_d, q_s$ are the quark charges. It is clear
that for basic nonet the ratio between $n \bar n$ and $s \bar s$ transition
form factors is just the ratio between charge factors. In case of radial
excitation nonet the wave function can change sign and the difference
between non-strange quark mass ($m_n=0.350$ GeV) and strange quark mass
($m_s=0.500$ GeV) affect  different behaviour of the transition form
factors.

Partial width for $\eta(M)$ decaying into $\gamma \gamma$ is given by
\begin{equation}
\Gamma_{\eta(M)\to\gamma\gamma} =
\frac\pi 4\alpha^2M^3
(F^{(1)}_{\eta(M)\to\gamma\gamma})^2\;, \label{g}
\end{equation}
where $\alpha=1/137$ and
\begin{eqnarray}
F^{(1)}_{\eta(1295)}&=&\cos\phi'\;\left[\cos\theta'\;
F^{(1)}_{n\bar n}-
\sin\theta'\;F^{(1)}_{s\bar s}\right] ,
\\
F^{(1)}_{\eta(1440)}&=&\cos\phi\left[\sin\theta\;
F^{(1)}_{n\bar n}+
\cos\theta\;F^{(1)}_{s\bar s}\right] \;.
\label{etawfsa}
\end{eqnarray}

\section{ $J/\Psi \to \gamma \eta(M)$ decay}

The radiative $J/\Psi$ decay is a source of additional information about
quark/gluon content of  $\eta$ mesons. The radiative $J/\Psi$ decay
is dominated by the annihilation of $c \bar c$
quarks, with the production of
photon and two gluons which form $q \bar q$ mesons. The amplitude for
such a process is proportional to the convolution of  two wave functions
$A \sim \Psi_{J/\Psi \to \gamma gg} \otimes \Psi_{meson}$, and
 corresponding branching ratio is
$BR = \mid A \mid^2 \times phase \,volume$.
So  the square of the production
amplitude is proportional to the probability of the glueball component
$R$ in the meson.  For $\eta(M)$ meson we have
\begin{equation}
R=\frac{BR(J/\Psi \to \gamma \eta(M))}{(M_{J/\Psi}^2 - M^2)^3} \; .
\end{equation}
Let us compare  probabilities $R$ for $\eta'$ and $\eta(1440)$.
Following \cite{PDG}
we use $$BR(J/\Psi \to \gamma \eta')= 4.3 \times
 10^{-3}$$
and   $$BR(J/\Psi \to \gamma \eta(1440))= 3.0 \times
 10^{-3}.$$ The calculations show that for
$\eta'$ and $\eta(1440)$ the values of $R$ are close
to each other. The result of Ref.\cite{AMN} shows that
$\eta'$ contains $(10 \pm 10)\%$ of the glueball component. It means that
we can expect the same amount of glueball for $\eta(1440)$.

The $\eta(1295)$ meson  is not seen in radiative $J/\Psi$ decay. The
 assumption  that
\begin{eqnarray}
   BR(J/\Psi \to \gamma \eta(1295)) = 0  \label{BR1295}
\end{eqnarray}
allows us to  estimate the
mixing angle between non-strange and strange $q \bar q$
component in $\eta(1295)$ and $\eta(1440)$. Since there is no glueball
admixture in $\eta(1295$ and $\sin\phi'=0$, the orthogonality
condition (\ref{ortcon}) gives us the same mixing angles for  $\eta(1295)$ and
 $\eta(1440)$: $\theta=\theta'$. As a result, we come to the following
mixing scheme:
\begin{eqnarray}
\Psi_{\eta(1295)}=\cos\theta\;\psi_{n\bar n}-
\sin\theta\;\psi_{s\bar s} ,
\end{eqnarray}
\begin{eqnarray}
\Psi_{\eta(1440)}=\cos\phi\left[\sin\theta\;\psi_{n\bar n}+
\cos\theta\;\psi_{s\bar s}\right]+\sin\phi\;\psi_G\ .
\label{etawfs1}
\end{eqnarray}

Then transition form factors are equal to
\begin{eqnarray}
F^{(1)}_{\eta(1295)}&=&\cos\theta\;
F^{(1)}_{n\bar n}-
\sin\theta\;F^{(1)}_{s\bar s} \ ,
\\
F^{(1)}_{\eta(1440)}&=&\cos\phi\left[\sin\theta\;
F^{(1)}_{n\bar n}+
\cos\theta\;F^{(1)}_{s\bar s}\right]\ .
\label{etawfsaa}
\end{eqnarray}

 The assumption (\ref{BR1295}) also
implies that $\eta(1295)$ is close to the SU(3) octet because the
 glueball is close to the SU(3) singlet. In fact there is no SU(3) flavour
symmetry, and the suppression parameter $\lambda$ for the production
   of $s \bar s$ pair is as follows \cite{lambda1,lambda2}:
\begin{eqnarray}
   \lambda=0.6 \pm 0.2\ .
\end{eqnarray}
Let us introduce flavour non-symmetrical singlet and octet states:
\begin{eqnarray}
\Psi_1=\frac{1}{\sqrt{3}}(\Psi_{u \bar u}
 + \Psi_{d \bar d} + \sqrt{\lambda} \Psi_{s \bar s}) \ ,\\
\Psi_8=\frac{1}{\sqrt{2+4/\lambda}}
(\Psi_{u \bar u} + \Psi_{d \bar d} - \frac{2}{\sqrt{\lambda}}
\Psi_{ s \bar s}) \ .
\end{eqnarray}
Glueball may turn into $\Psi_1$ while the transition "$glueball \rightarrow
\Psi_8$" is forbidden.
 This means that $\Psi_{\eta(1295)}= \Psi_8$ and therefore
\begin{eqnarray}
\cos \theta=
\sqrt{\frac{\lambda}{2+\lambda}} \ .
\label{xx}
\end{eqnarray}

\section{Results}

Since the two-photon width of $\eta(1440)$ was found in $KK \pi$
decay channel  we can  estimate  this branching
ratio. The BES collaboration claimed \cite{BES98}
that $\eta \pi \pi$ decay is weak:
$10-20\%$ of $K \bar K$.  Similar ratios have been found in the
by Mark III \cite{Mark3} and DM2 \cite{DM2}.

On the opposite, the data on $p \bar p \to \eta(1440) \sigma$ from
\cite{CBAR} give the following ratio:
\begin{eqnarray}
BR(\eta(1440) \to KK\pi)/BR(\eta(1440) \to \eta \pi\pi) =
0.61 \pm 0.19\;.     \label{add1}
\end{eqnarray}
A large discrepancy between these two samples of data
can be explained by the fact that in $p \bar p$ annihilation the phase
space available in the $\sigma$ amplitude is limited; furthermore, its
amplitude goes to zero for low $\sigma$ mass. The result is a strong
suppression of the upper side of $\eta(1440)$. As was shown in
\cite{BES98}, the suppression increases in a factor 3  between 1440 MeV
and 1465 MeV which makes weaker the process $\eta(1440) \to K^*K$ in $p
   \bar p$ annihilation.

Using eq.(\ref{add1}), we assume the following limits
for the $\eta(1295)$ and $\eta(1440)$ two-photon widths:
\begin{eqnarray}
160  \; \mathrm{eV}\; < \Gamma_{\gamma\gamma}(\eta(1440))
\;<\; 300  \; \mathrm{eV}\;,    \label{limit1}           \\
\Gamma_{\gamma\gamma}(\eta(1295))
\;<\; 66  \; \mathrm{eV}\;.    \label{limit2}
\end{eqnarray}

\begin{figure}[h!]
\begin{tabular}{c}

\mbox{\epsfig{file=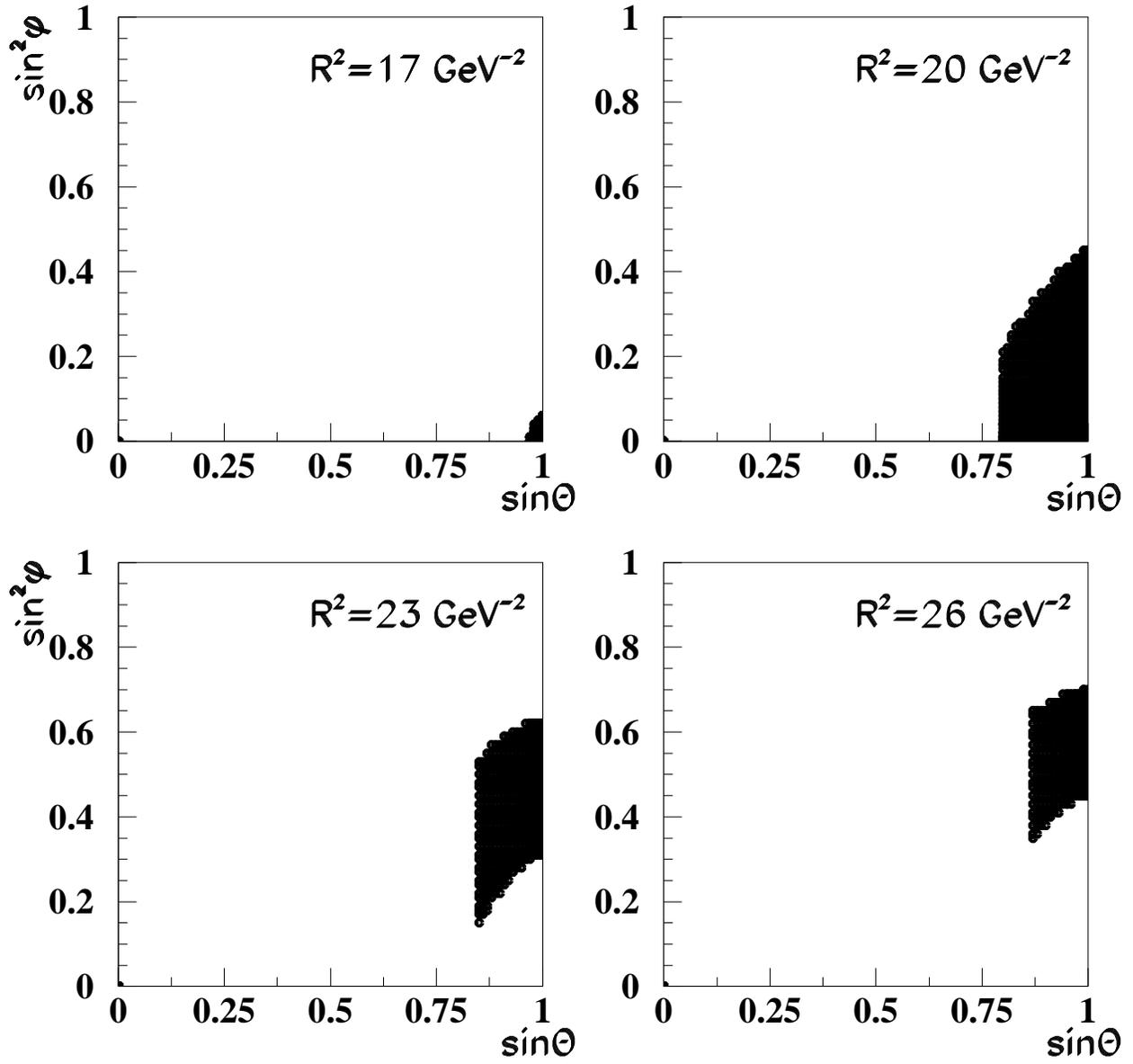,width=1.\textwidth,height=1.\textwidth}
}
\end{tabular}
    \caption{\small \it
Allowed ($\theta,\phi$)-region following from the
conditions (\ref{limit1}) and (\ref{limit2}) at differnent
mean radius squared of $\eta(1440)$.}
\label{fig-plot}
\end{figure}

Fig. \ref{fig-plot} shows us the allowed region for mixing angle
   $\theta$ and glueball admixture $\phi$ which comes from the
constraints (\ref{limit1}) and (\ref{limit2}). Since the transition
form factors depend strongly on $R^2$, the allowed ($\theta,\phi$)
region depends on $R^2$. For larger $R^2$ it is possible that
$\eta(1440)$ has  more glueball admixture. The condition
(\ref{limit2}) restricts the amount of $n \bar n$ component in
$\eta(1295)$ as well,
thus $\eta(1440)$ is expected to be mainly $n \bar n$ state for
different $R^2$. One can expect that the first radial excitation state
is larger than the basic one, and
it seems natural to consider mean radius squared  in the region
$R^2 \sim 15-20 $ GeV$^{-2}$. Together with
additional limits  (\ref{xx}) for  mixing
angle, we have:
\begin{eqnarray}
0.84 \;<\; \sin \theta \;<\;0.91\;,
\end{eqnarray}
and we come to the following limits for the glueball admixture in $\eta(1440)$:
\begin{eqnarray}
0\; < \sin^2 \phi \; <\;0.40 \ .
\end{eqnarray}
Let us note that this result does not contradict the  naive estimation which
follows from the comparison of radiative $\eta'$ and $\eta(1440)$ decays.

At first glance
the obtained results, which comes from the two-photon decay and radiative
$J/\psi$ decay,  contradict the experimental branching
rations for $\eta(1295$ and $\eta(1440)$. The $\eta(1440)$  has a strong
$K \bar K \pi$ decay mode. For example, BES Collaboration found \cite{BES98} the
branching ratios to $ KK^*$, $KK_0$ and $\eta \pi\pi$ in the ratios 1.92: 0.83:
1. Meanwhile, all recent observations of $\eta(1295)$ have been done in $\eta
\pi\pi$ decay mode. If $\eta(1295)$ has a large $s \bar s$ component,  one
can expect the dominant decay mode with strange mesons.

\begin{figure}
\begin{tabular}{c}

\mbox{\epsfig{file=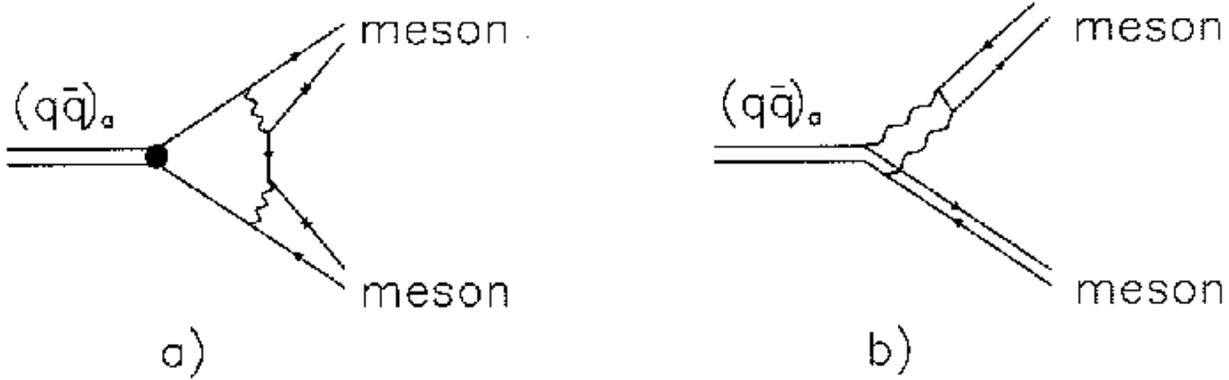,width=1.\textwidth}
}
\end{tabular}
\caption{\small \it Diagrams for the decay of the  $q\bar q$-state
into two mesons.}
\label{FIG-DECAY}
\end{figure}

Let us discuss in detail the meson decay mechanism. Experimental data
indicate that the decays of $\eta(1295)$ and $\eta(1440)$ are the cascade
reactions so first the two-meson state is produced. In our case it can be
$ KK^*$, $KK_0$ or $a_0(980) \pi$. The quark combinatorial rules \cite{combin}
may be applied to the calculation of corresponding decay couplings.
 There exist two types of transitions "$q\bar q \rightarrow two \; mesons$"
shown in Fig. \ref{FIG-DECAY}.  The type of process
represented by the diagram of Fig. \ref{FIG-DECAY}a
is leading one, according to the rules of $1/N$ expansion. The coupling constants
for $\eta$ meson decaying into two strange mesons ($ KK^*$ or $KK_0$)
in the leading order of $1/N$ expansion are:
\begin{eqnarray}
g_{\eta(1440) \rightarrow KK} = g^L (\sqrt \lambda\sin\theta +
 \sqrt2\cos\theta)\;,   \nonumber
  \\
g_{\eta(1295) \rightarrow KK} = g^L (\sqrt \lambda\cos\theta -
 \sqrt2\sin\theta)\;,
\end{eqnarray}
 where the parameter  $g^L$ hides the unknown dynamics of the decay.
With $\sin\theta = 0.85$ and $\lambda=0.6$ the flavour factor gives the
suppression of the $\eta(1295)$ coupling constant by a factor 2 (factor 4 in
branching ratio) as compared with $\eta(1440)$ coupling constant. Another
suppression comes from the kinematical factor, since phase volume for $
KK^*$ and $KK_0$ in $\eta(1295)$ decay is much smaller than in $\eta(1440)$
decay. These arguments can explain the fact why no $KK\pi$ is observed in
$\eta(1295)$ decay.

Finally, we need to comment the fact that $\eta(1440)$, with
large non-strange $q \bar q$ component, has a higher mass than
 $\eta(1295)$ state which is $s \bar s$-dominant one.
This can be the result of the $q \bar q \to gluons$
transitions which can be different for singlet and octet states, thus
shifting the mass of meson. Let us remind
that similar situation takes place in the ground nonet where   $\eta'$ is
heavier than other nonet members (so called $U_A(1)$ problem \cite{u1}).
There are different models which explain this phenomenon, for example,
instanton approach \cite{instanton}, and   it is possible
 that the instanton-induced forces can also
give such effect for  $\eta(1295)$ and  $\eta(1440)$.

\section{Conclusion}

Recent results on
$\eta(1295)$ and  $\eta(1440)$ two-photon widths from L3
\cite{L3} together with the information from radiative $J/\Psi$ decay
support the hypothesis that these resonances are
members of the first radial excitation $2 ^1S_0$ $q \bar q$ nonet.
We have
determined the  $\eta(1295)/ \eta(1440)$ mixing angle and admixture
of the glueball component using the calculation scheme developed in \cite{AANM}.
We argue that $\eta(1440)$ is dominantly
non-strange $q \bar q$ state, with possible admixture of glueball
component $(20\pm 20)\%$ while  $\eta(1295)$ is mainly $s \bar s$
state with small gluonium component.
 These estimations are essentially based on the fact that
there is just one resonance state near 1440 MeV, so
further analysis of the experimental data is needed to solve this problem.

\section{Acknowledgements}

I thank V.V Anisovich, D. Bugg  and I. Vodopianov
for useful discussions and comments.
Special thank M. Kienzle-Focacci for invitation to visit Geneva University
where most of the work was done.

 The work was partly supported by the RFBR grant 01-02-17861.


\end{document}